\documentclass[structabstract]{aa}  
\usepackage{graphicx}
\usepackage{txfonts}
\begin{document}
   \title{The galactic unclassified B[e] star HD\,50138}

   \subtitle{III. The short-term line profile variability of its photospheric lines\thanks{Based on observations made with the Mercator Telescope, operated on the island of La Palma by the Flemish Community, at the Spanish Observatorio del Roque de los Muchachos of the Instituto de Astrof\'isica de Canarias.}\thanks{Reduced spectra are only available in electronic form at the CDS via anonymous ftp to cdsarc.u-strasbg.fr (130.79.128.5) or via http://cdsweb.u-strasbg.fr/cgi-bin/qcat?J/A+A/}}
   \author{M. Borges Fernandes
          \inst{1}
          \and
          M. Kraus
          \inst{2}
          \and
	  D.~H. Nickeler
	  \inst{2}
          \and
          P. De Cat
	  \inst{3}
          \and
          P. Lampens
	  \inst{3}
          \and
          C. B. Pereira
          \inst{1}
          \and
          M. E. Oksala
          \inst{2}
          }

   \institute{Observat\'orio Nacional, Rua General Jos\'e Cristino 77, 20921-400 S\~ao Cristov\~ao, Rio de Janeiro, Brazil\\
              \email{borges@on.br, claudio@on.br}
         \and
             Astronomick\'y \'ustav, Akademie v\v{e}d \v{C}esk\'e republiky, Fri\v{c}ova 298, 251\,65 Ond\v{r}ejov, Czech Republic\\
             \email{kraus@sunstel.asu.cas.cz, nickeler@asu.cas.cz, oksala@sunstel.asu.cas.cz}
         \and
             Royal Observatory of Belgium, Ringlaan 3, B-1180 Brussels, Belgium\\
             \email{Peter.DeCat@oma.be, patricia.lampens@oma.be}
             }

   \date{Received ; accepted }

 
  \abstract
   {HD\,50138 presents the B[e] phenomenon, but its nature is not clear yet. This star is known to present spectral variations, which have been associated with outbursts and shell phases.}
   {We analyze the line profile variability of HD\,50138 and its possible origin, which provide possible hints to its evolutionary stage, so far said to be close to the end of (or slightly beyond) the main sequence.}
   {New high-resolution spectra of HD\,50138 obtained with the HERMES spectrograph over
several nights (five of them consecutively) were analyzed, allowing us to confirm short-term line profile variability.}
   {Our new data show short-term variations in the photospheric lines. On the other hand, purely circumstellar lines (such as [O\,{\sc i}] lines) do not show such rapid variability. The rotational velocity of HD\,50138, $\varv_{\rm rot} = 90.3\pm 4.3$\,km\,s$^{-1}$, and the rotation period, $P= 3.64\pm 1.16$\,d, were derived from the He\,{\sc i}\,$\lambda 4026$ photospheric line. Based on the moment method, we confirm that the origin of this short-term line profile variability is not stellar spots, and it may be caused by pulsations. In addition, we show that macroturbulence may affect the profiles of photospheric lines, as is seen for B supergiants.}
   {The location of HD\,50138 at the end of (or slightly beyond) the main sequence, the newly detected presence of line profile variability resembling pulsating stars, and macroturbulence make this star a fascinating object. }


   \maketitle
%

\section{Introduction}

The group of B[e] stars displays, in their optical spectrum, strong Balmer emission lines and emission from 
permitted and forbidden transitions of neutral and singly ionized metals. In addition, B[e] stars show circumstellar dust, which is responsible for the excess emission 
in the near and mid-infrared. Especially in the Galaxy, many B[e] stars are 
still unclassified with respect to their evolutionary phase, mainly due to
unknown or uncertain stellar parameters and especially distances. In addition, 
they are often embedded in dense circumstellar material. Consequently, their 
optical spectra are strongly polluted with circumstellar emission, while 
unpolluted photospheric lines, which are needed for a proper stellar classification, are 
rare or absent.

The star \object{HD\,50138} (V743\,Mon, MWC\,158) is one of these galactic unclassified B[e] stars 
(Lamers et al. \cite{Lamers}), although strong effort has been undertaken to determine its 
stellar parameters. Borges Fernandes et al. (\cite{Paper1}, hereafter Paper\,I) reclassified it as a 
B6-7\,III-V star based on high-resolution optical spectra and discussed several 
evolutionary scenarios. It seems that HD\,50138 might be a star that is close to the end of 
its main-sequence evolution. This indicates that this star may represent a link between B[e] and Be stars. Shell phases and outbursts of this star have been reported, based on line profile variabilities
(Merrill \cite{Merrill}; Merrill \& Burwell \cite{MerrillBurwell}; Hutsem\'{e}kers \cite{Hutsemekers}; Andrillat \& Houziaux \cite{Andrillat}; Pogodin \cite{Pogodin}). And indeed, photometric observations have shown a drop in $UBV$ magnitudes by about 0.2\,mag, which is interpreted as an outburst around 1978/79. In Paper\,I, it was also suggested that a new shell phase took place prior to 2007. 

It is also important to point out that HD\,50138 is often discussed in the framework of Herbig Ae/Be stars (see Paper\,I). Based on the absence of any nebulosity around it, as well as the absence of a close-by star-forming region, it could be ``an isolated" HAeB[e] star. However, such a classification is unlikely because of the presence of shell phases and outbursts, which are typical of more evolved B[e] stars.

The presence of a circumstellar disk has been inferred from polarimetric and spectro-polarimetric observations
that revealed an intrinsic polarization in agreement with a non-spherically 
symmetric distribution of the circumstellar material (Vaidya et al. \cite{Vaidya}; Bjorkman et al. \cite{Bjorkman}; Oudmaijer \& Drew \cite{Oudmaijer}; Harrington \& Kuhn \cite{Harrington}). Recently, Borges Fernandes et al. (\cite{Paper2}, hereafter Paper\,II), based on interferometric measurements, confirmed the presence of a gaseous and dusty disk around HD\,50138. The orientation on the sky plane and a proper inclination angle of $56\degr\pm 4\degr$ were also derived.  
From both spectroscopic and interferometric observations, the presence
of a companion is unlikely, although it cannot be definitely excluded yet.

To better understand the nature of its spectral variability, we re-observed HD\,50138 during several nights (most of which were consecutive) in 2011. Thus, we present in Sect. 2 our observations and data reduction. In Sect. 3, we present our results from these high-quality spectra and in Sect. 4 and 5, we discuss the new facts about HD\,50138 and present our conclusions, respectively.


\section{Observations}

High-quality optical spectra were obtained with the High Efficiency and Resolution Mercator Echelle Spectrograph (HERMES) attached to the 1.2-m Mercator telescope at the Roque de los Muchachos Observatory on La Palma (Canary Islands, Spain). HERMES is a fiber-fed spectrograph (Raskin et al. \cite{Raskin}) and our data were obtained in the high-resolution mode (HRF), which provides a spectral resolution of $R\simeq 85\,000$ and a field of view of 2.5\,arcsec$^2$. The wavelength coverage extends from 3770\,\AA \ to 9000\,\AA. For a proper reduction we used the automated data reduction pipeline, HermesDRS, which includes the barycentric correction.

The observations were obtained during eight nights in the period 2011 February 19 to 27. Five of the nights were consecutive. During each night, for optimum cosmic ray elimination, we took two consecutive spectra, which were checked for any line variability, and in the case of no change, were then combined for a better signal-to-noise ratio ($S/N$). Details on the resulting spectra are summarized in 
Table\,1.

\begin{table}[ht]
\label{obslog}
\begin{center}
  \caption{Journal of available spectra.} 
\tabcolsep 1.5 pt
\begin{tabular}{ccccc}
\\
\hline\hline
\noalign{\smallskip}
$Date$  & BJD-2455600 & UT      & $T_{\rm exp}$  & $S/N$ \\
      &   [d]         & [h]:[m] &  [s] & \\
\noalign{\smallskip} \hline \noalign{\smallskip}
2011-02-19 & 11.5249327 & 00:36 & 720(2)  &  220  \\
2011-02-20 & 12.5581295 & 01:24 & 900(2)  &  140  \\
2011-02-22 & 14.5001431 & 00:00 & 900(2)  &  145  \\
2011-02-22 & 15.4722788 & 23:20 & 600(2)  &  190  \\
2011-02-23 & 16.4782732 & 23:29 & 750(2)  &  240  \\
2011-02-24 & 17.4157888 & 21:59 & 825(2)  &  230  \\
2011-02-25 & 18.4739716 & 23:23 & 900(2)  &  230  \\
2011-02-27 & 20.4782182 & 23:29 & 1200(2)  &  220  \\
\noalign{\smallskip} \hline \noalign{\smallskip}
\end{tabular}
\tablefoot{The barycentric Julian date (BJD) and the universal time (UT) correspond to the middle of the combined exposures; number in brackets to the exposure times gives the total number of exposures; signal-to-noise ($S/N$) values refer to the combined spectra.}
\end{center}
\end{table}

\section{Results}\label{results}

\subsection{Line profile variability}\label{lpv}

\begin{figure*}[!tbh]
\centering   
\includegraphics[width=0.48\hsize]{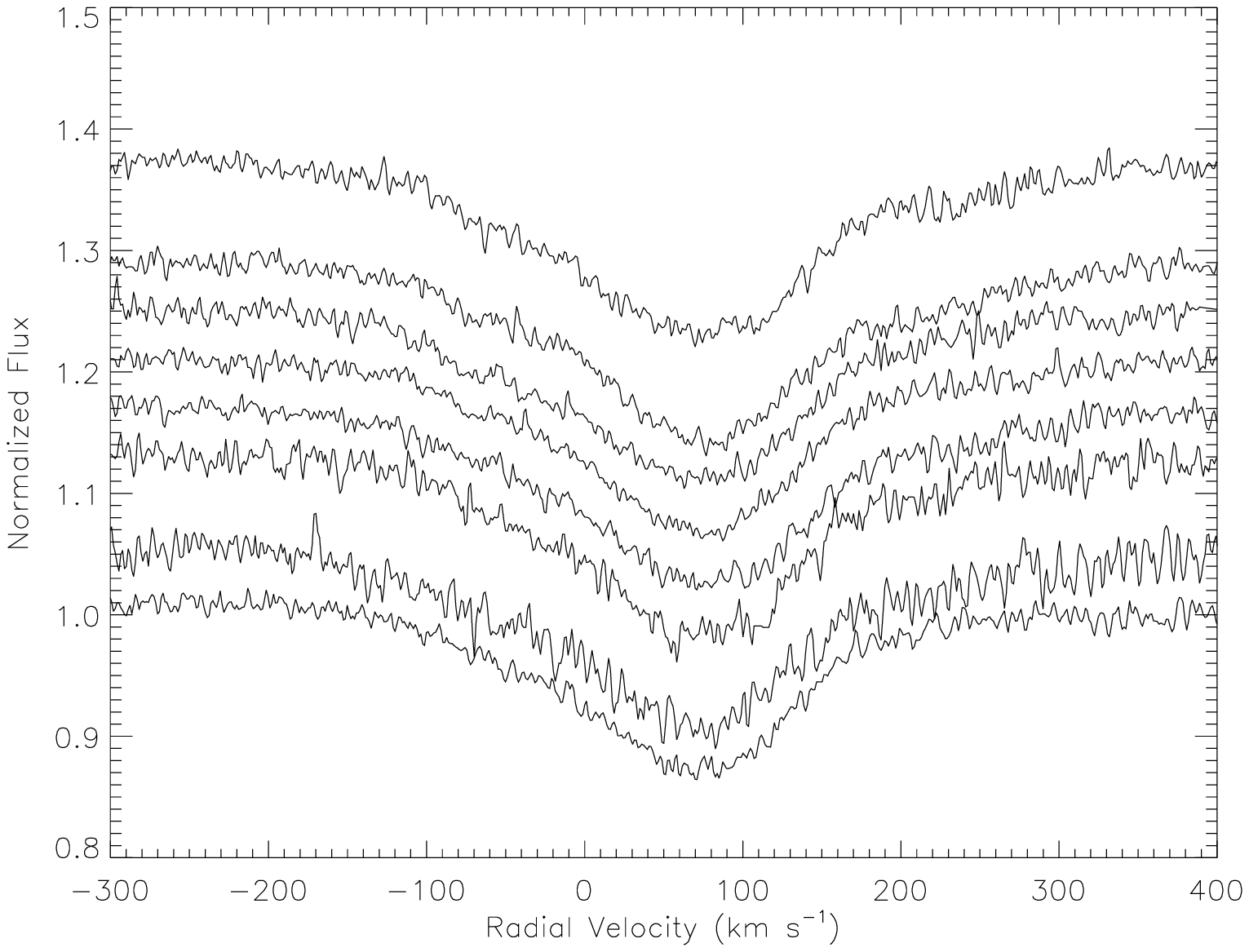}\qquad
\includegraphics[width=0.48\hsize]{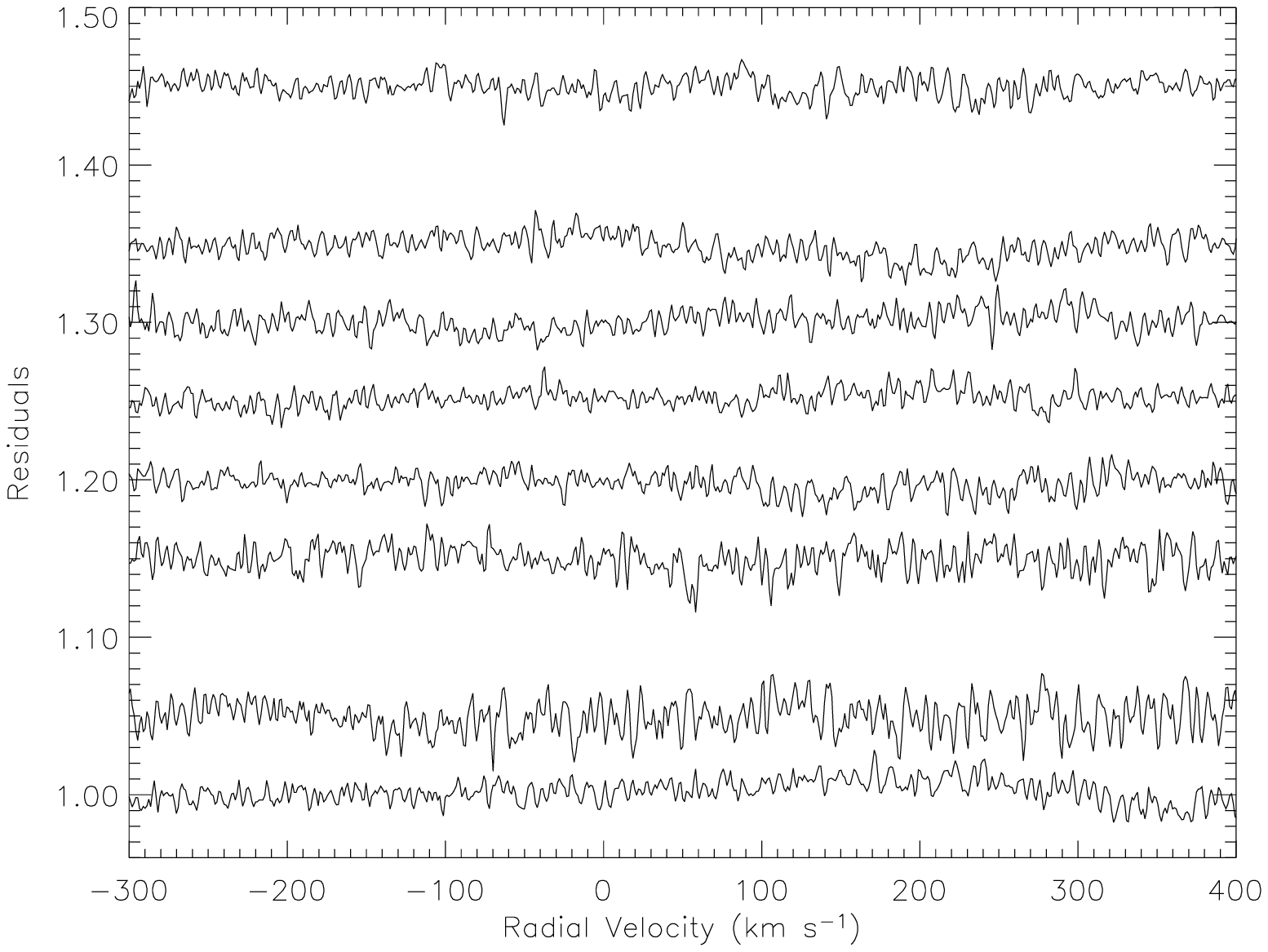}\qquad
\caption{He\,{\sc i}\,$\lambda 4026$ line. The left panel shows its time series and the right panel shows the night-to-night residuals. The time increases from bottom (2011-02-19) to top (2011-02-27), and for a better visual inspection of the spectra, they are slightly shifted along the vertical axis with offsets proportional to the interval between the observations, considering the barycentric Julian dates. }\label{HeI4026}
\end{figure*}

\begin{figure*}[!tbh]
\centering   
\includegraphics[width=0.48\hsize]{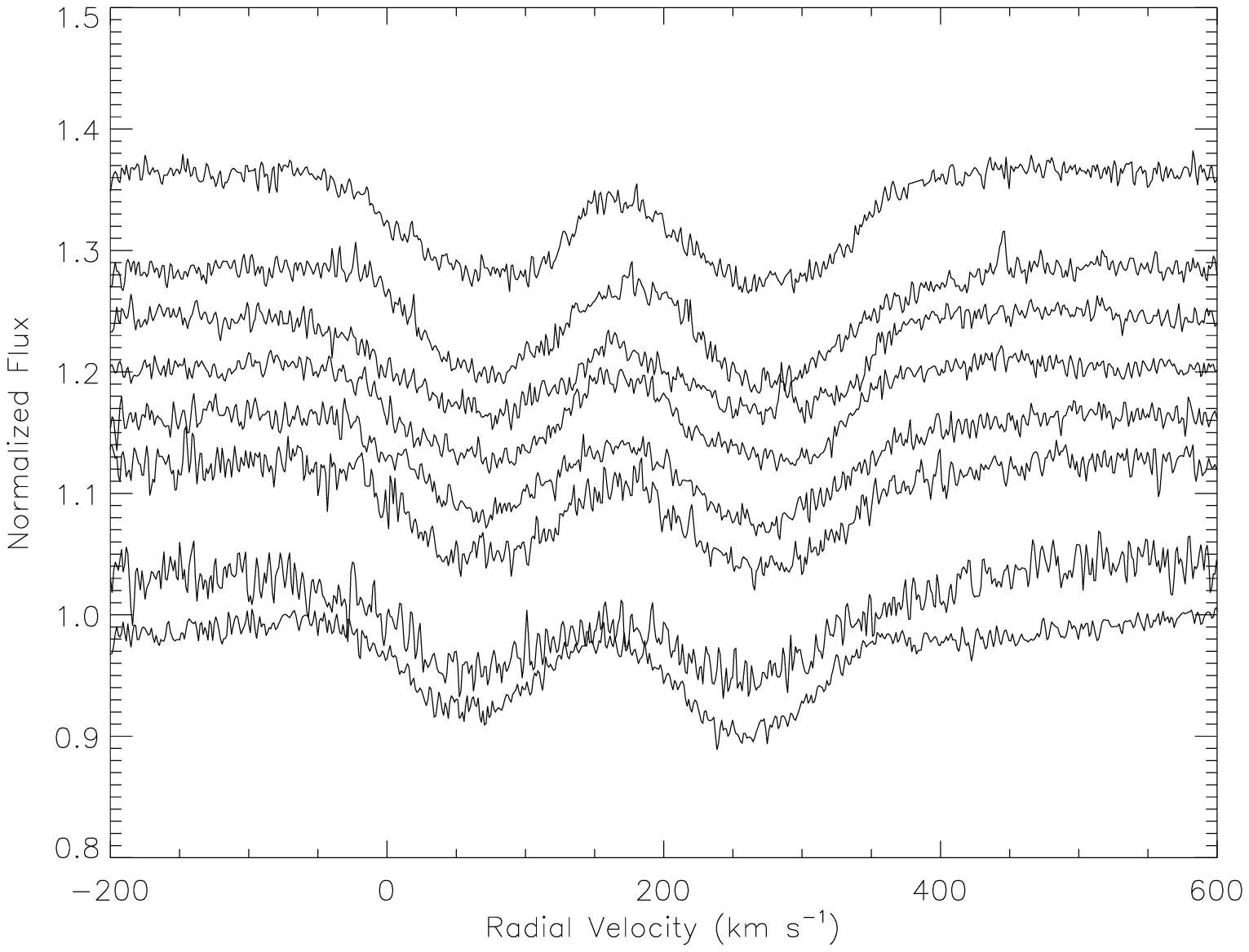}\qquad
\includegraphics[width=0.48\hsize]{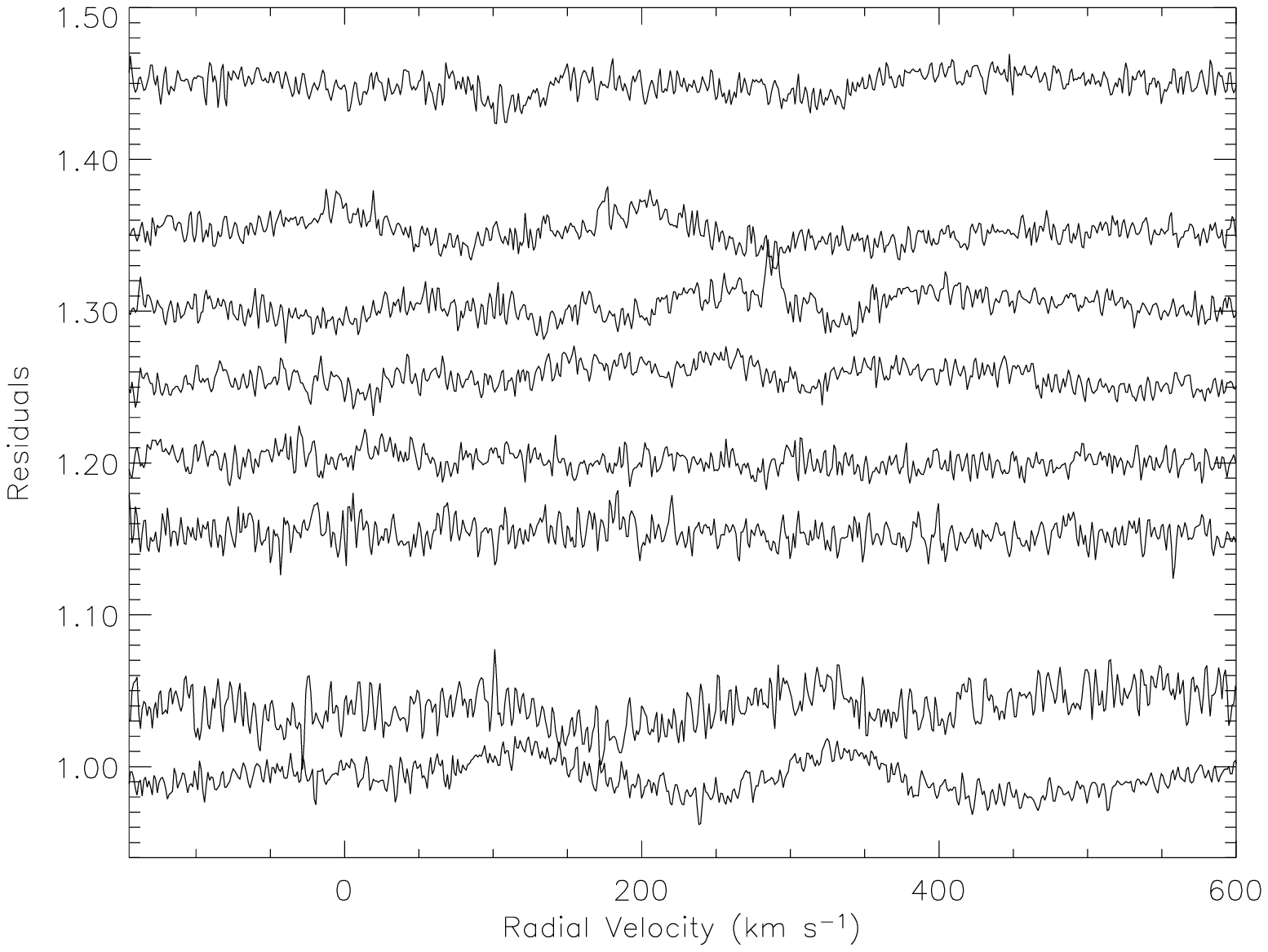}\qquad
\caption{As in Fig.\,1, but for Si\,{\sc ii}\,$\lambda \lambda 4128, 4131$ lines. The radial velocity was derived as a function of the wavelength of rest of the $\lambda 4128$ line.}\label{SiII4128}
\end{figure*}

The spectrum of HD\,50138, as presented in Paper\,I, is dominated by lines from neutral and singly ionized elements. It also presents a rich diversity of line profiles from emission to pure absorption lines, including several lines with a combination of absorption and emission components. 

Line profile variability has previously been reported in the literature and associated with shell phases and outbursts (Merrill \cite{Merrill}; Merrill \& Burwell \cite{MerrillBurwell}; Hutsem\'{e}kers \cite{Hutsemekers}; Andrillat \& Houziaux \cite{Andrillat}; Pogodin \cite{Pogodin}). Paper\,I reported long-term variability in the high-resolution spectra taken eight years apart. The variability seen in several lines was interpreted as a new shell phase, which took place prior to 2007.

\begin{figure*}[!tbh]
\centering   
\includegraphics[width=0.48\hsize]{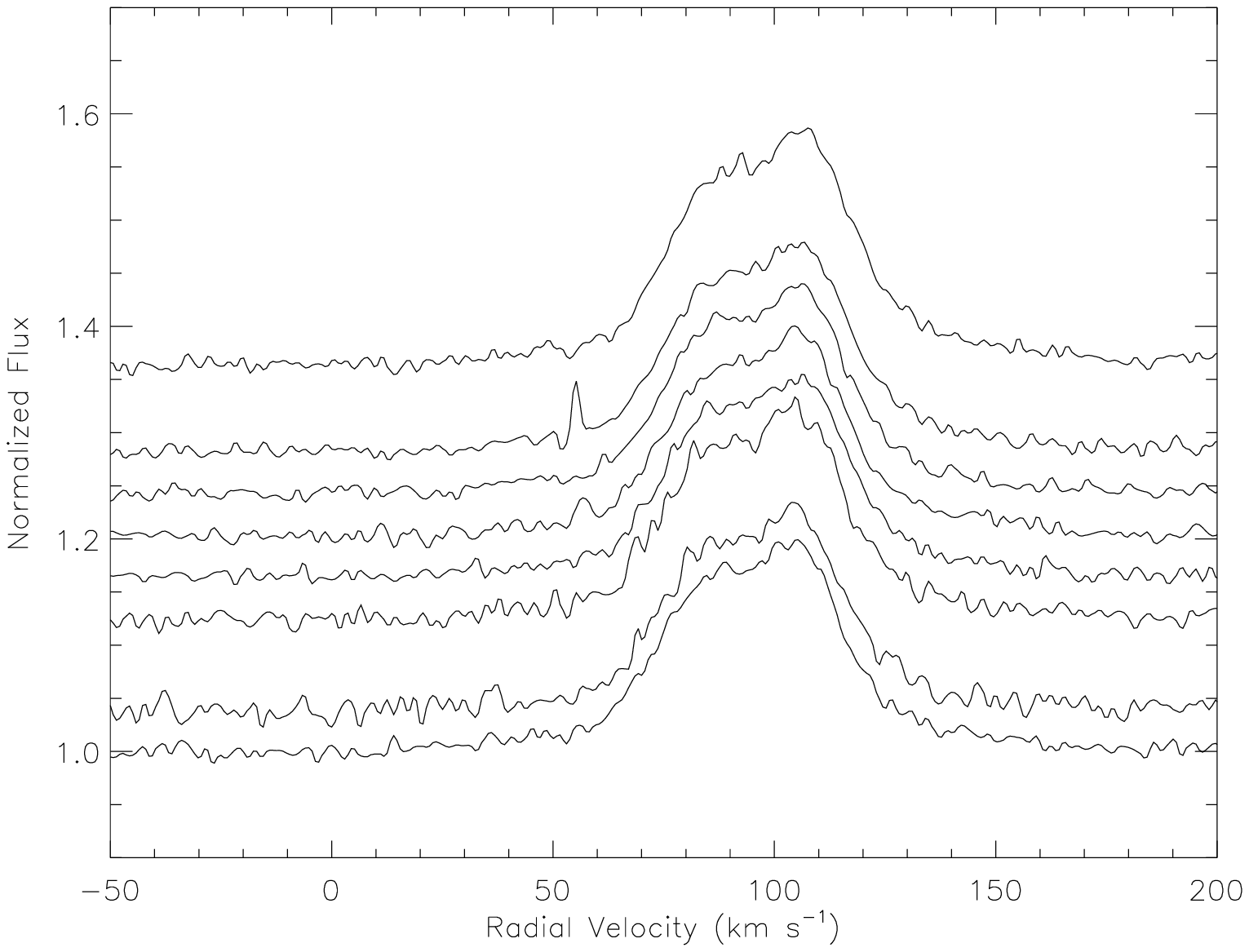}\qquad
\includegraphics[width=0.48\hsize]{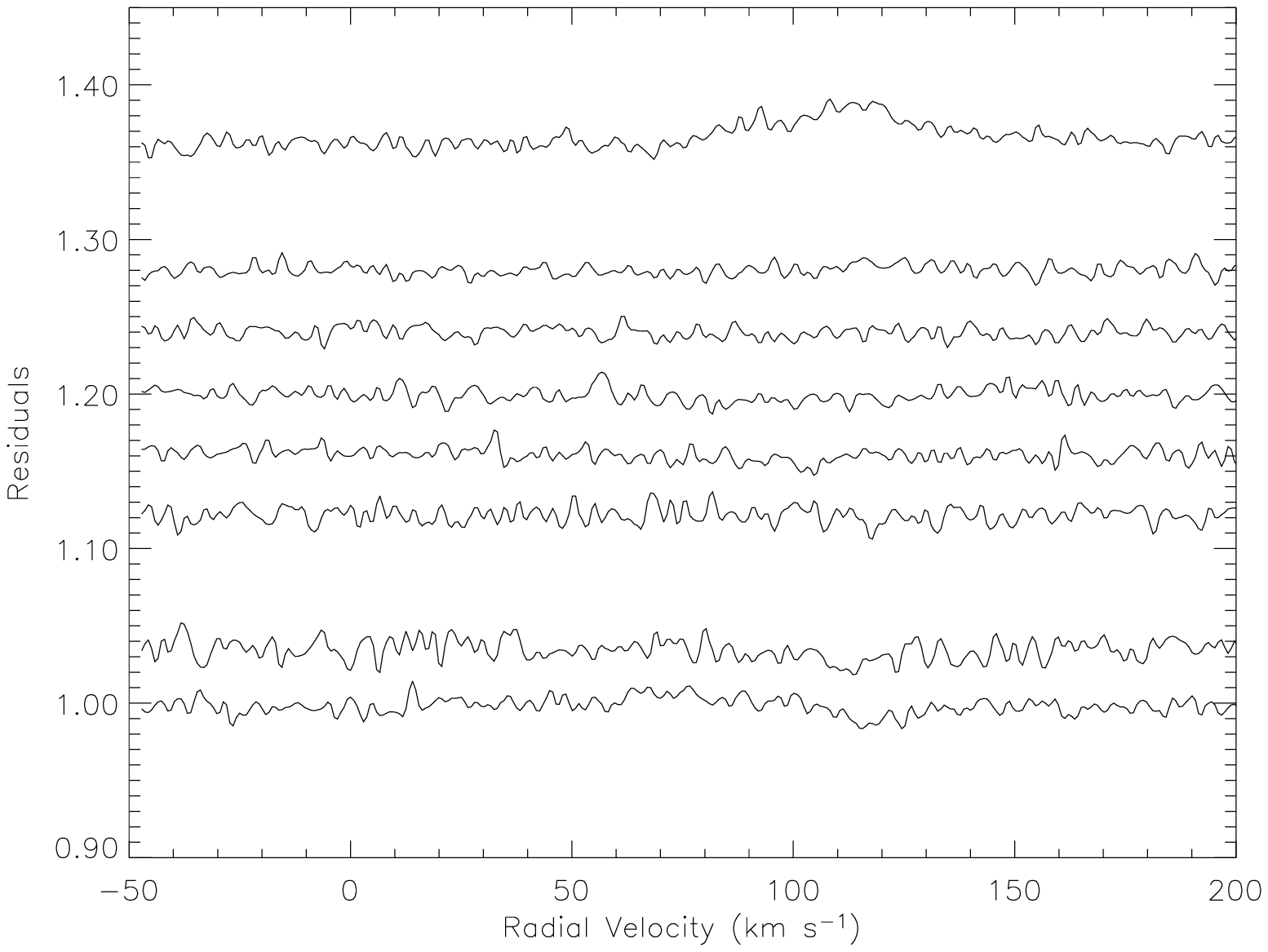}\qquad
\caption{As in Fig. 1, but for [O\,{\sc i}]\,$\lambda 6364$ line.}\label{OI6364}
\end{figure*}

\begin{figure*}[!tbh]
\centering   
\includegraphics[width=0.48\hsize]{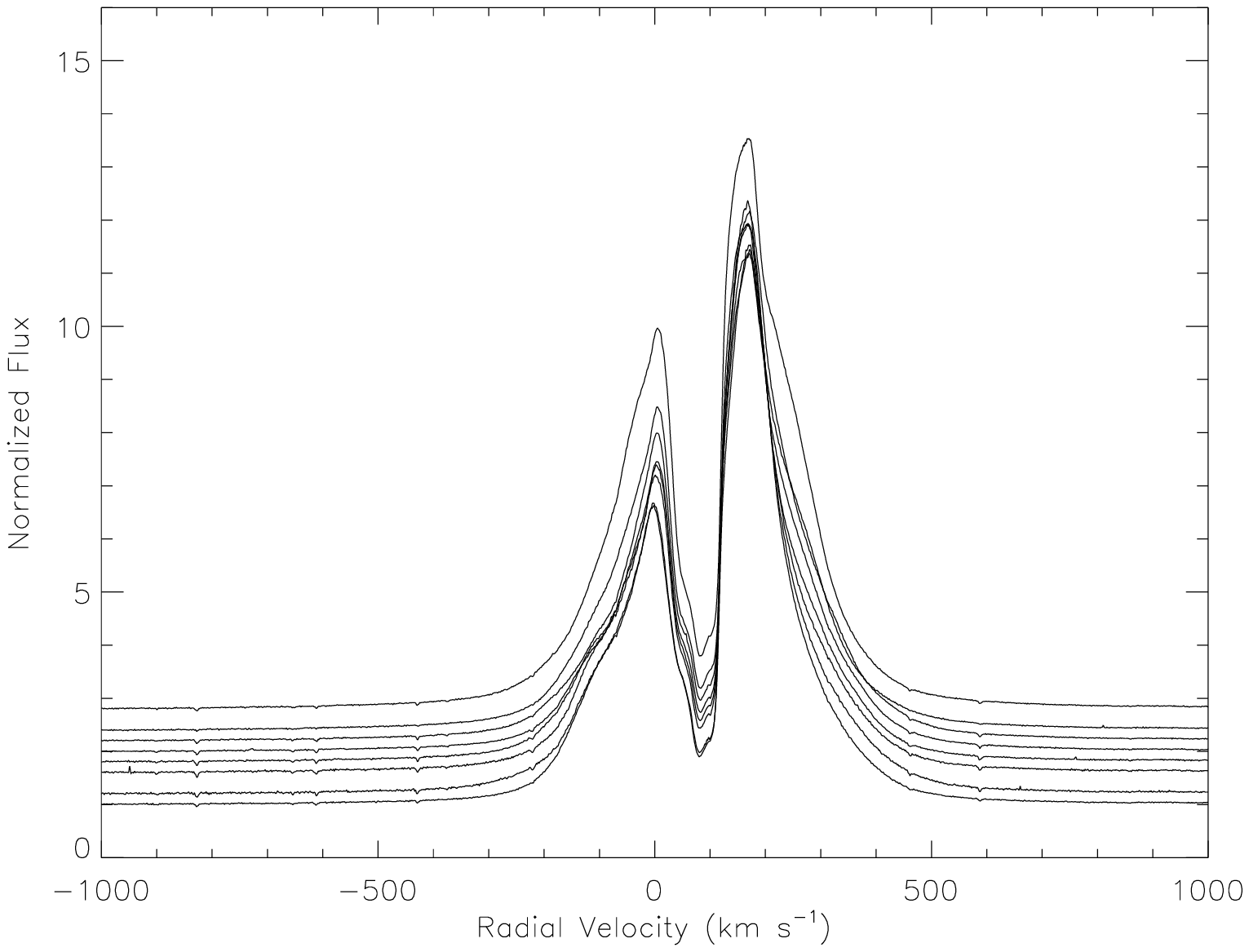}\qquad
\includegraphics[width=0.48\hsize]{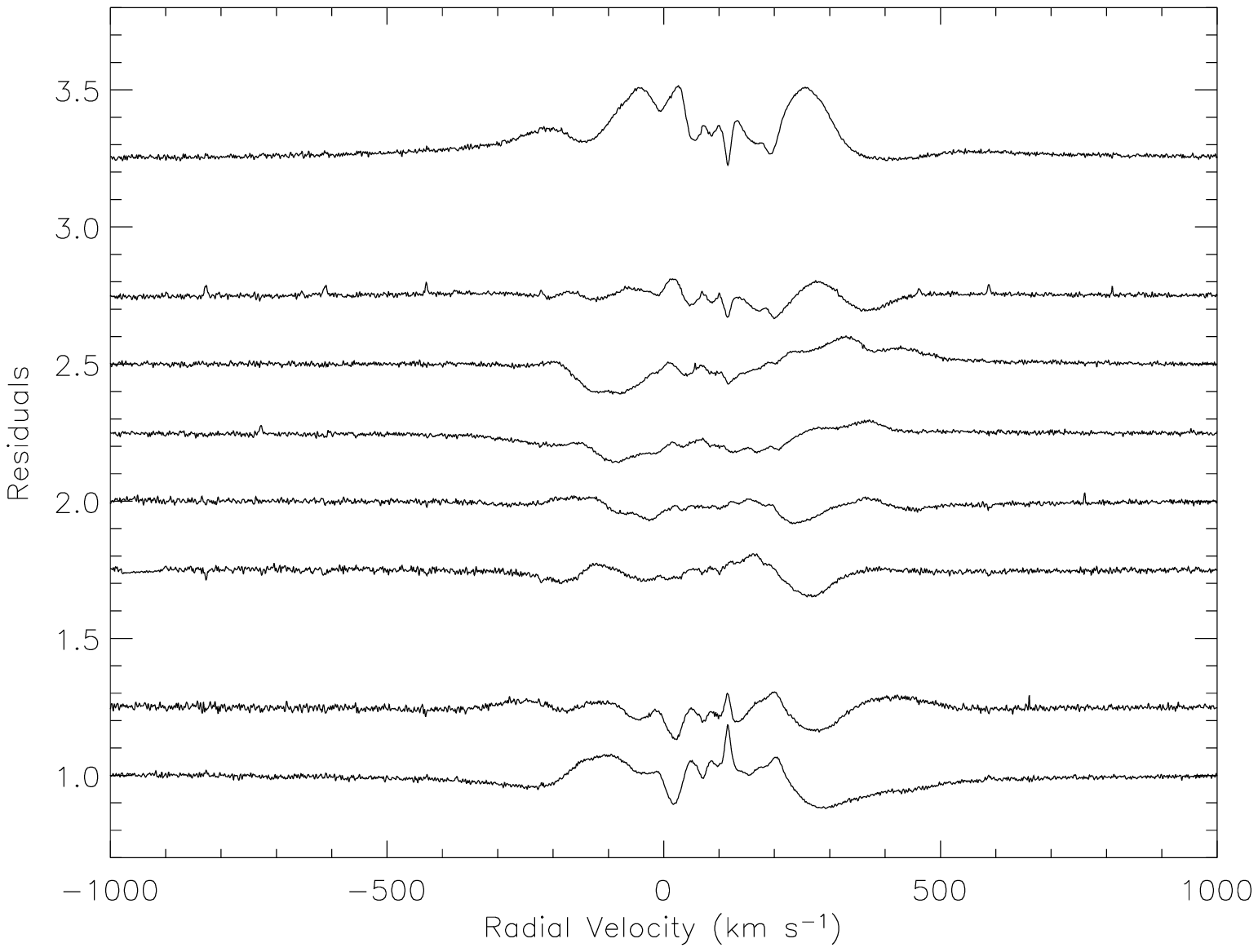}\qquad
\caption{As in Fig. 1, but for H$\alpha$\,$\lambda 6563$ line.}\label{Ha}
\end{figure*}

On the other hand, Pogodin (\cite{Pogodin}) also reported night-to-night variability in the line profiles, from spectra taken in four consecutive nights in March 1994, but covering small spectral regions. From our HERMES data taken during several consecutive nights with higher resolution and broader wavelength coverage, we confirm this short-term variability for a large sample of lines of different chemical elements. In Figs.\,\ref{HeI4026} through \ref{Ha}, we show some examples of the lines seen in our spectra, noting the different strengths of variability, which seem to be related to the formation region of these lines. 

The spectrum of HD\,50138 displays a few photospheric lines. 
In fact, the only purely photospheric lines seem to be He\,{\sc i}\,$\lambda 4026$ and Si\,{\sc ii}\,$\lambda \lambda 4128, 4131$. These lines, as can be seen by their residuals\footnote{All available spectra were added and a mean spectrum was obtained and then, the residuals were calculated as the ratio of the individual ones with the mean spectrum.} in Figs.\,\ref{HeI4026} and \ref{SiII4128}, do not display strong variability. These lines are known to be formed at deeper atmospheric layers. Other possible photospheric lines like He\,{\sc i}\,$\lambda$ 4009, 4124, and 4123 are hardly detected, and He\,{\sc i}\,$\lambda$ 4388 is blended in our spectra.

On the other hand, lines like the red doublet of Si\,{\sc ii}\,$\lambda\lambda 6347, 6371$ (Figs.\,\ref{SiII6347} and \ref{SiII6371}), He\,{\sc i} lines in the red portion of the spectrum, such as He\,{\sc i}\,$\lambda 6678$ (Fig.\,\ref{HeI6678}) and the core of the higher Balmer lines (Figs.\,\ref{H10} and \ref{H9}), present strong variations from night to night. These lines are formed in the upper layers of the stellar atmosphere or very close to the stellar surface, and are extremely sensitive to an enhancement or reduction of the local density due to any disturbance in their formation regions.

\onlfig{3}{
\begin{figure*}[!tbh] 
\includegraphics[width=0.48\hsize]{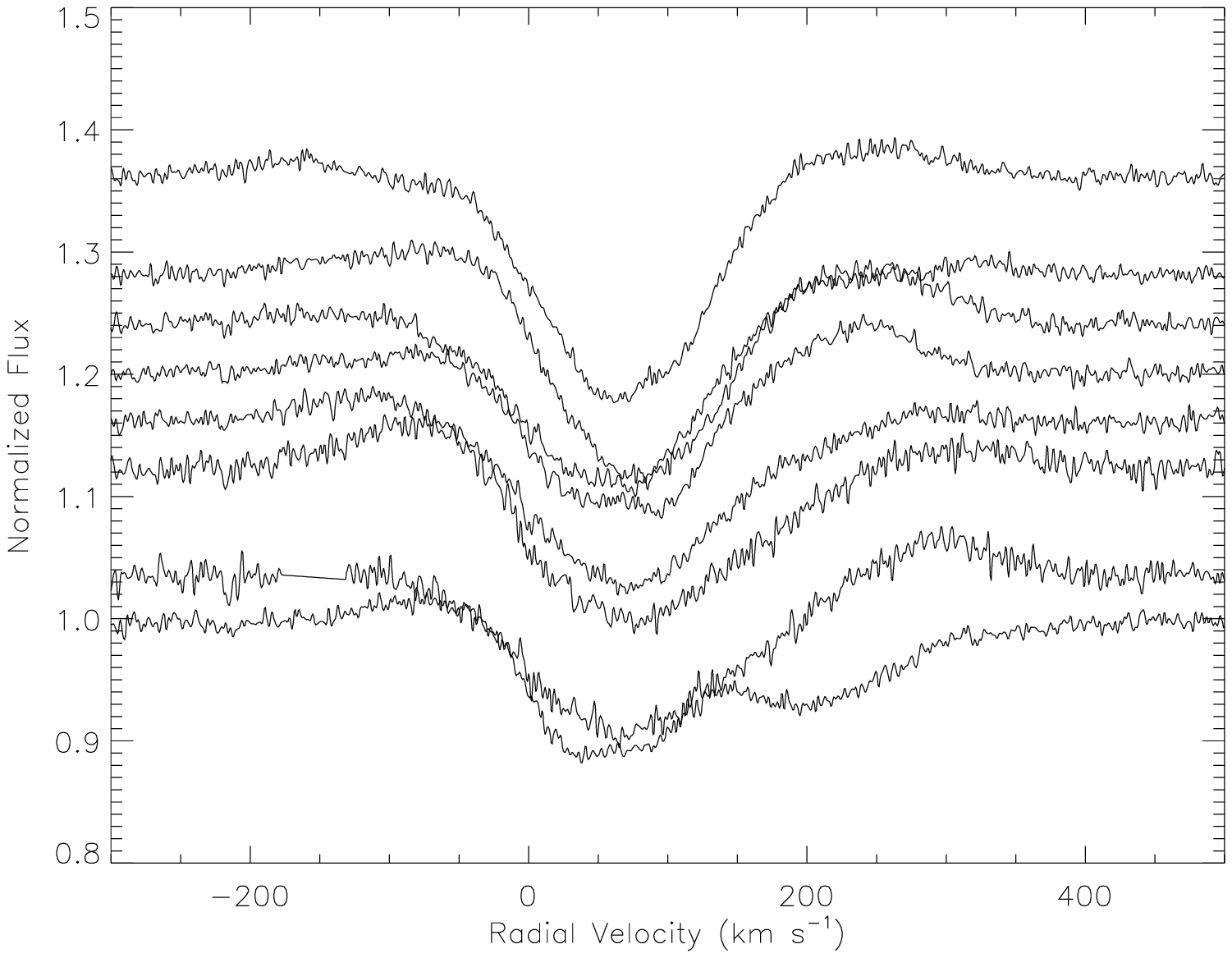}\qquad
\includegraphics[width=0.48\hsize]{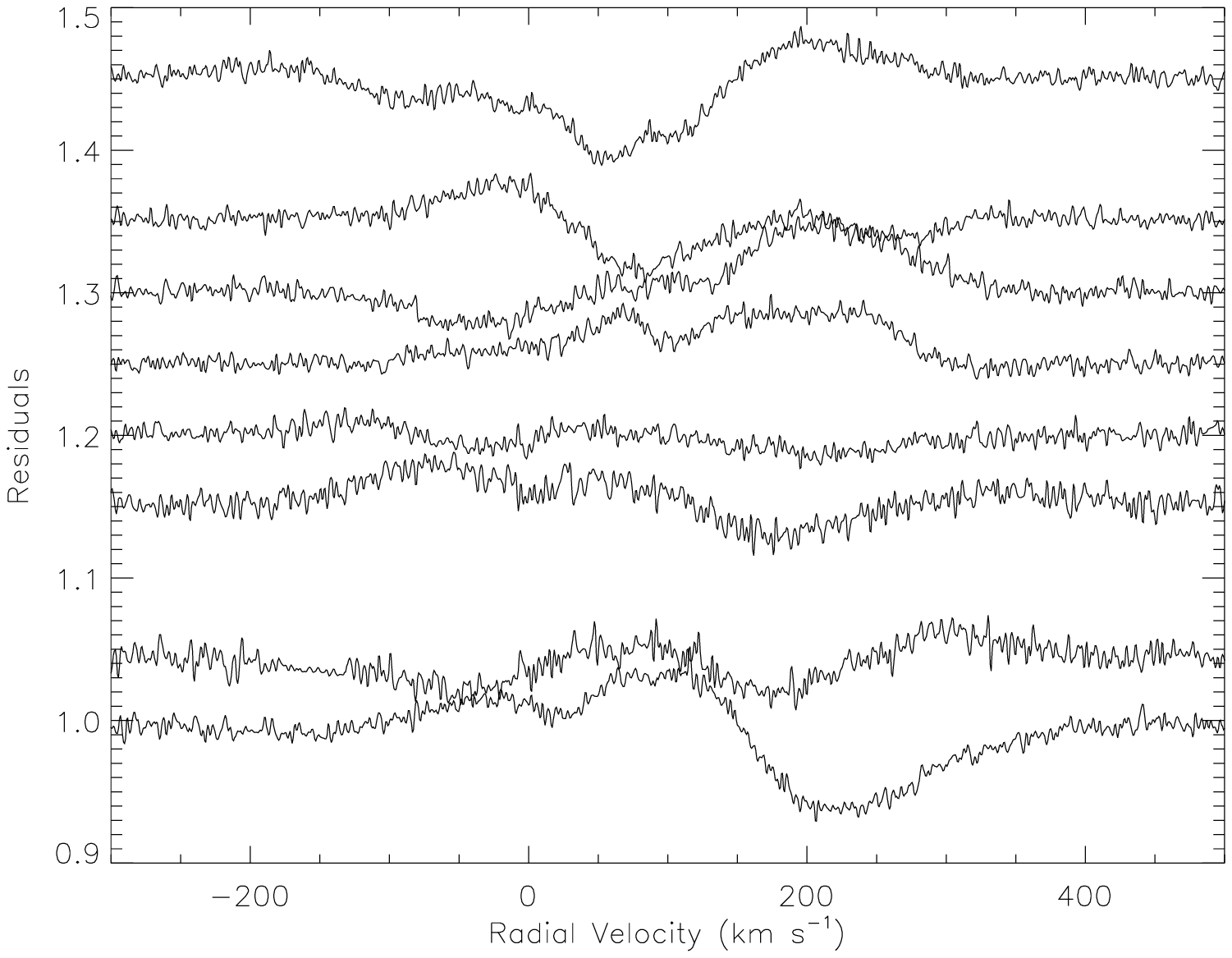}\qquad
\caption{As in Fig.\,1, but for Si\,{\sc ii}\,$\lambda 6347$ line. }
\label{SiII6347}
\end{figure*}
}

\onlfig{4}{
\begin{figure*}[!tbh]
\includegraphics[width=0.48\hsize]{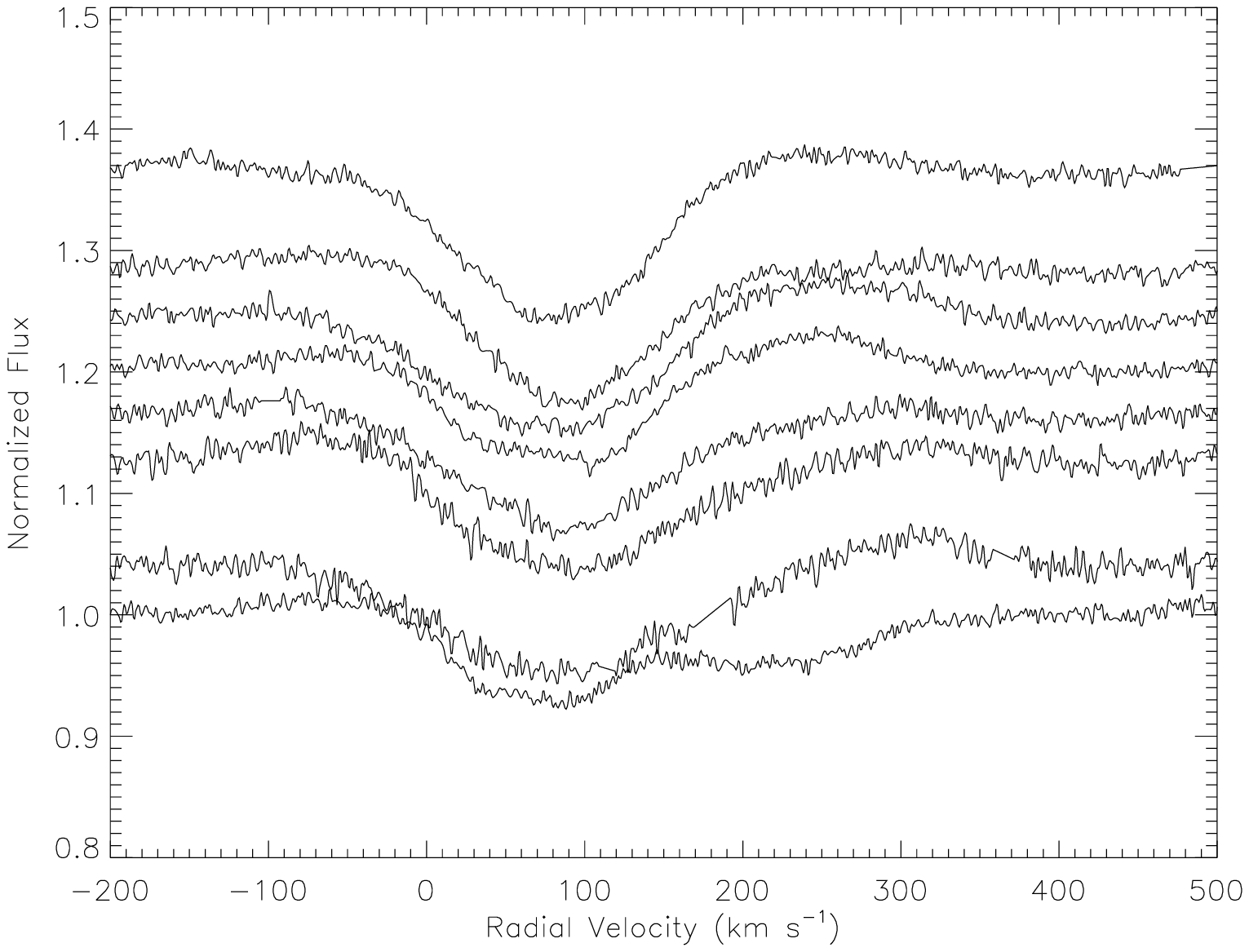}\qquad
\includegraphics[width=0.48\hsize]{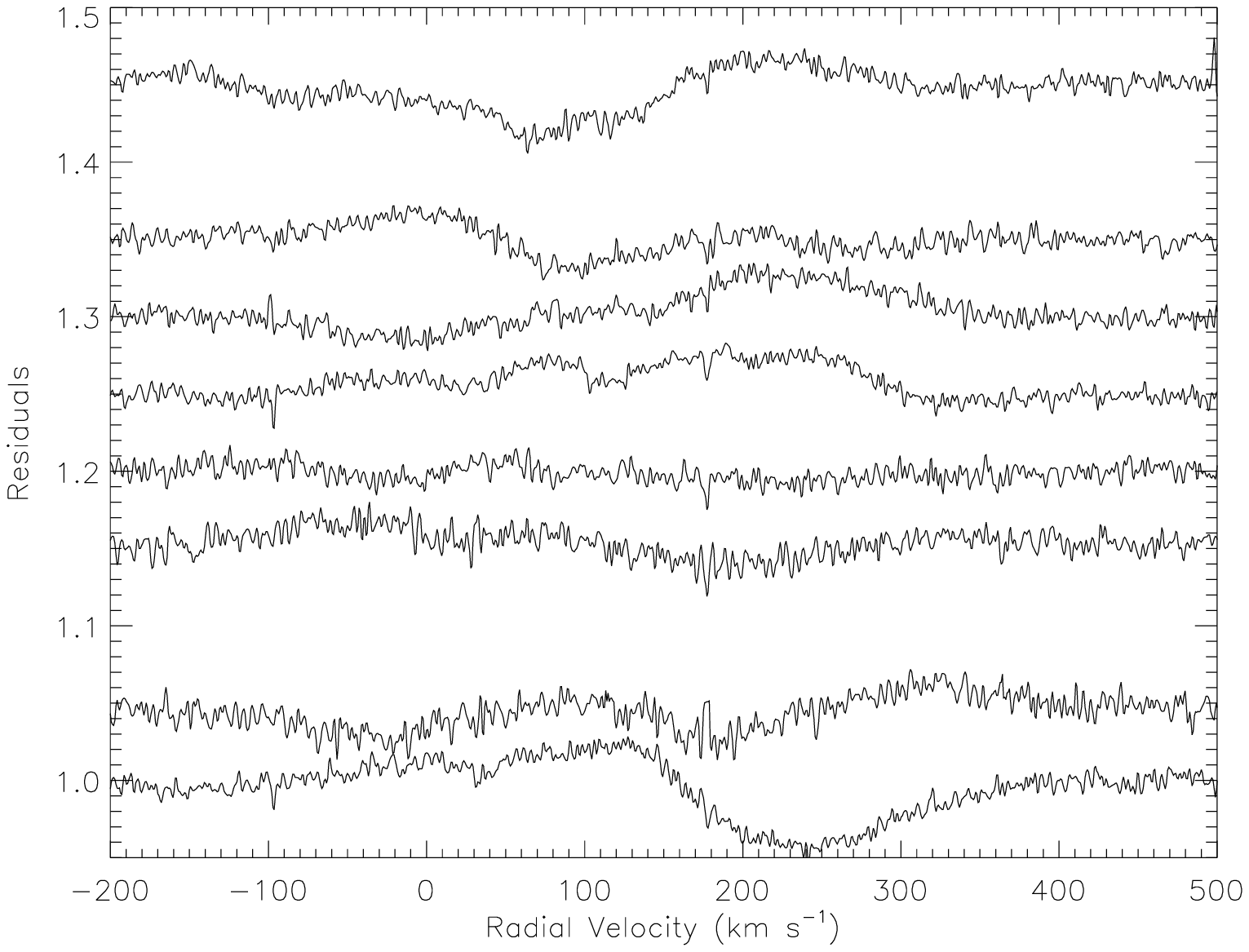}\qquad 
\caption{As in Fig.\,1, but for Si\,{\sc ii}\,$\lambda 6371$ line.}
\label{SiII6371}
\end{figure*}
}


The emission lines of circumstellar origin, such as the forbidden lines [O\,{\sc i}]\,$\lambda\lambda 6300, 6364$, are quite stable. This can be seen in the Fig.\,\ref{OI6364} for the [O\,{\sc i}]\,$\lambda 6364$ line and its residuals. Since no telluric correction was performed for our data, we do not present the [O\,{\sc i}]\,$\lambda 6300$ line, which is polluted by telluric lines. 

However, the same stability is not seen for H$_\alpha$, as can be seen from its residuals (Fig.\,\ref{Ha}). H$_\alpha$ is usually formed in the outer photosphere and lower (ionized) wind or disk region (1-5 $R_*$), while the [O\,{\sc i}] lines are formed at much greater distances ($>$ 100 $R_*$), where the disk material is already predominantely neutral (Kraus et al. \cite{Kraus2007}; \cite{Kraus2010}). 

\onlfig{5}{
\begin{figure*}[!tbh]
\includegraphics[width=0.48\hsize]{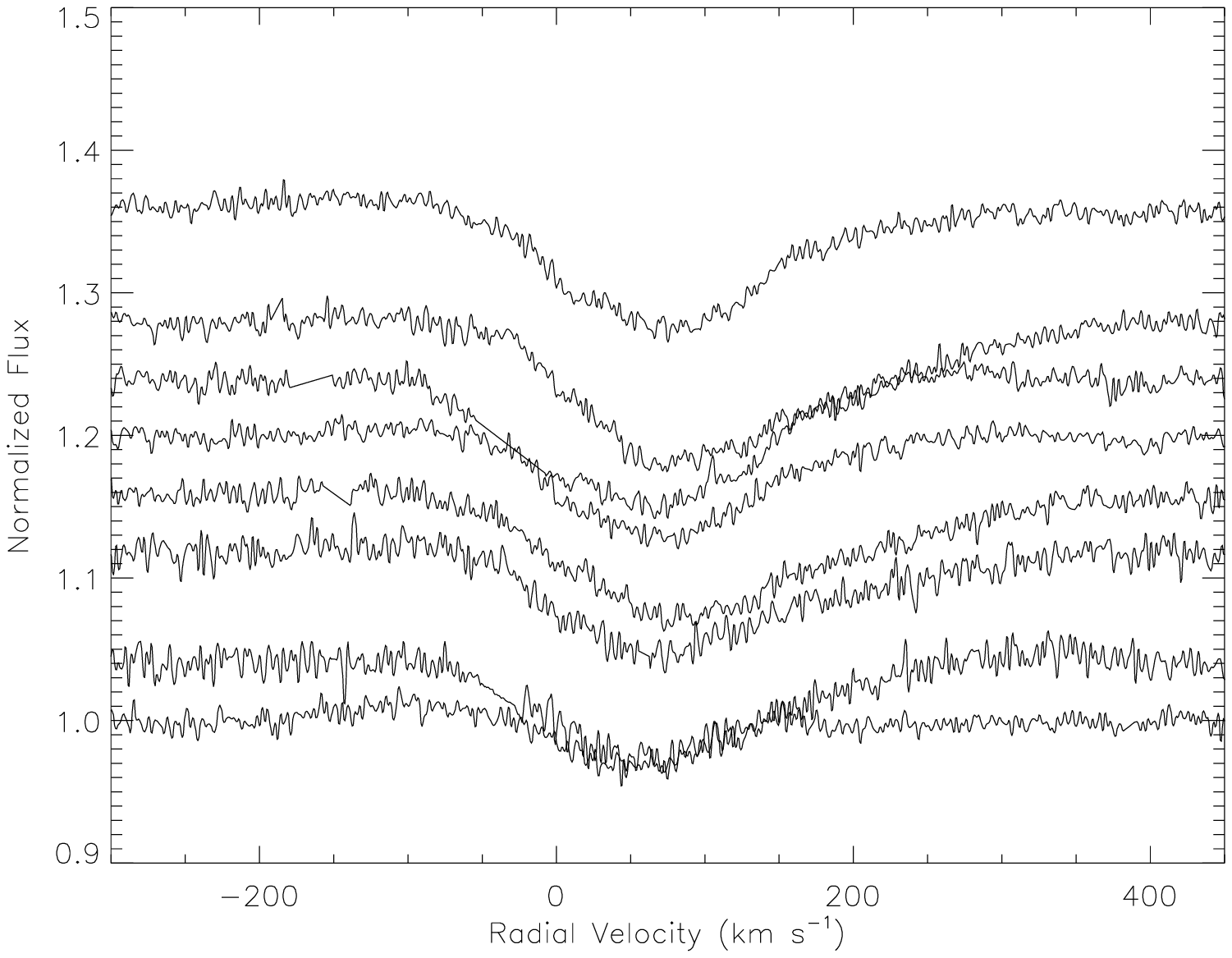}\qquad
\includegraphics[width=0.48\hsize]{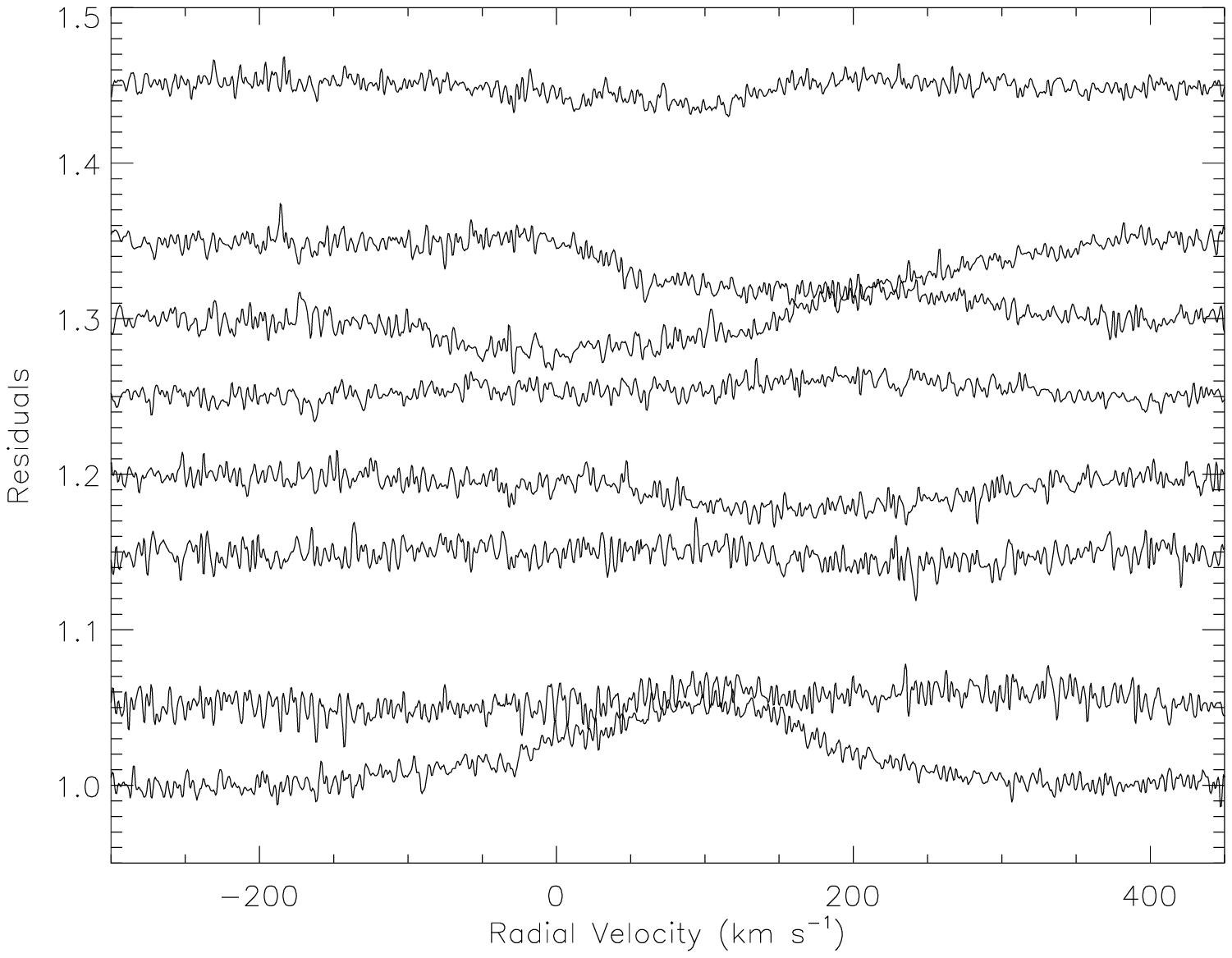}\qquad 
\caption{As in Fig.\,1, but for He\,{\sc i}\,$\lambda 6678$ line.}
\label{HeI6678}
\end{figure*}
}

\onlfig{6}{
\begin{figure*}[!tbh]
\includegraphics[width=0.48\hsize]{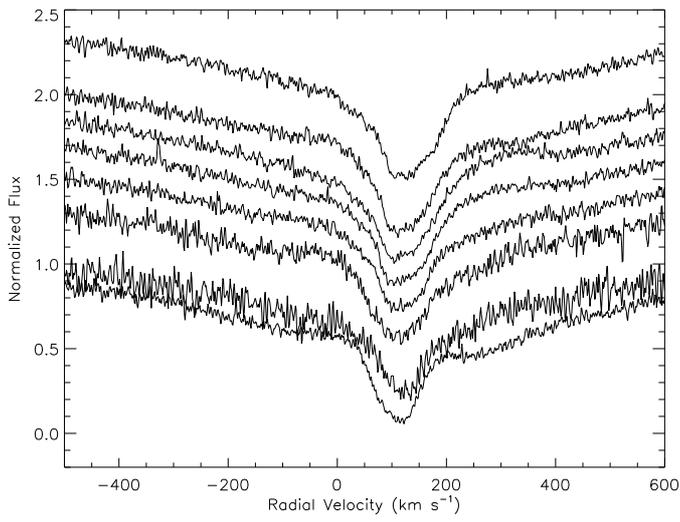}\qquad
\includegraphics[width=0.48\hsize]{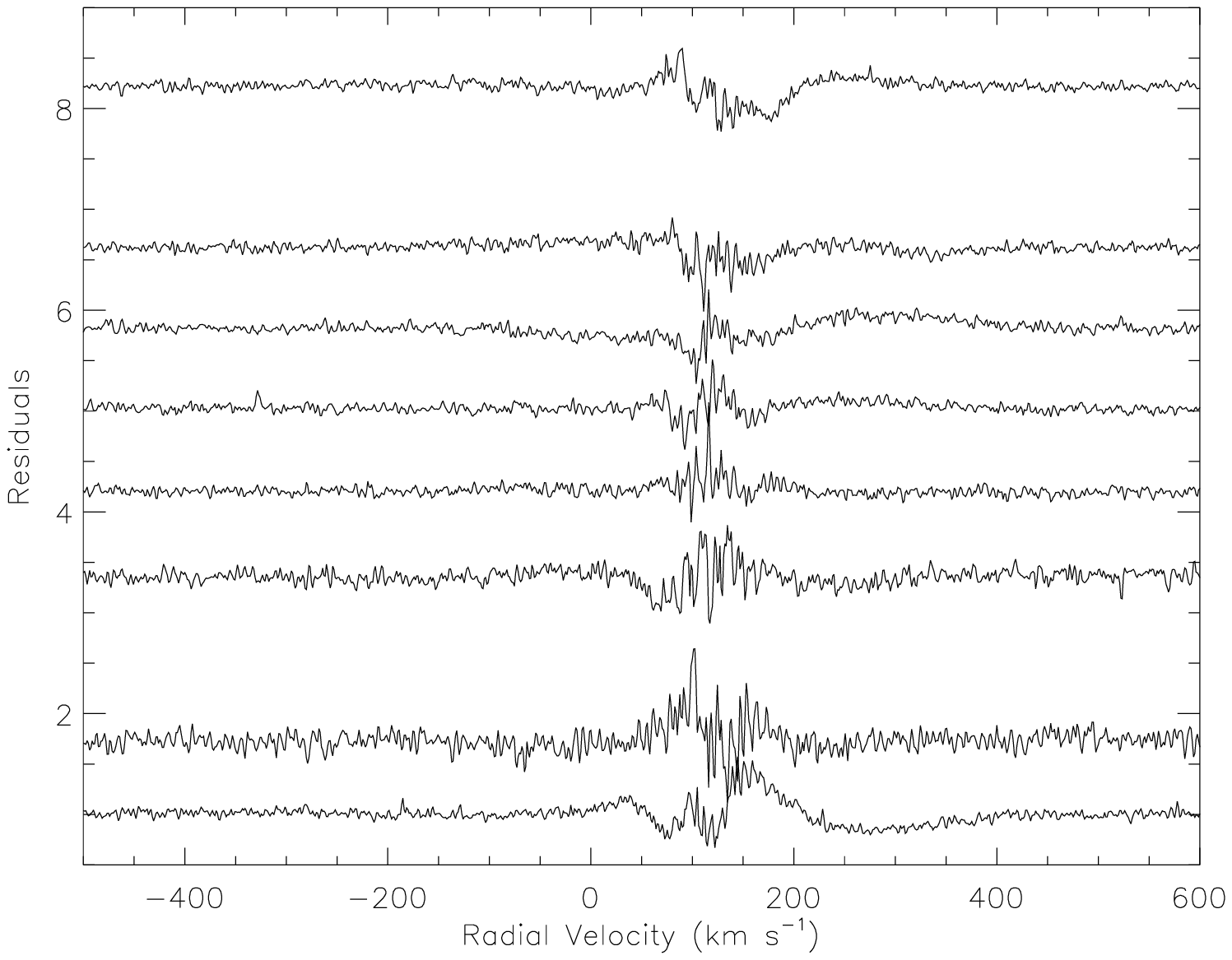}\qquad 
\caption{As in Fig.\,1, but for H\,10\,$\lambda 3798$ line.}
\label{H10}
\end{figure*}
}

\onlfig{7}{
\begin{figure*}[!tbh]
\includegraphics[width=0.48\hsize]{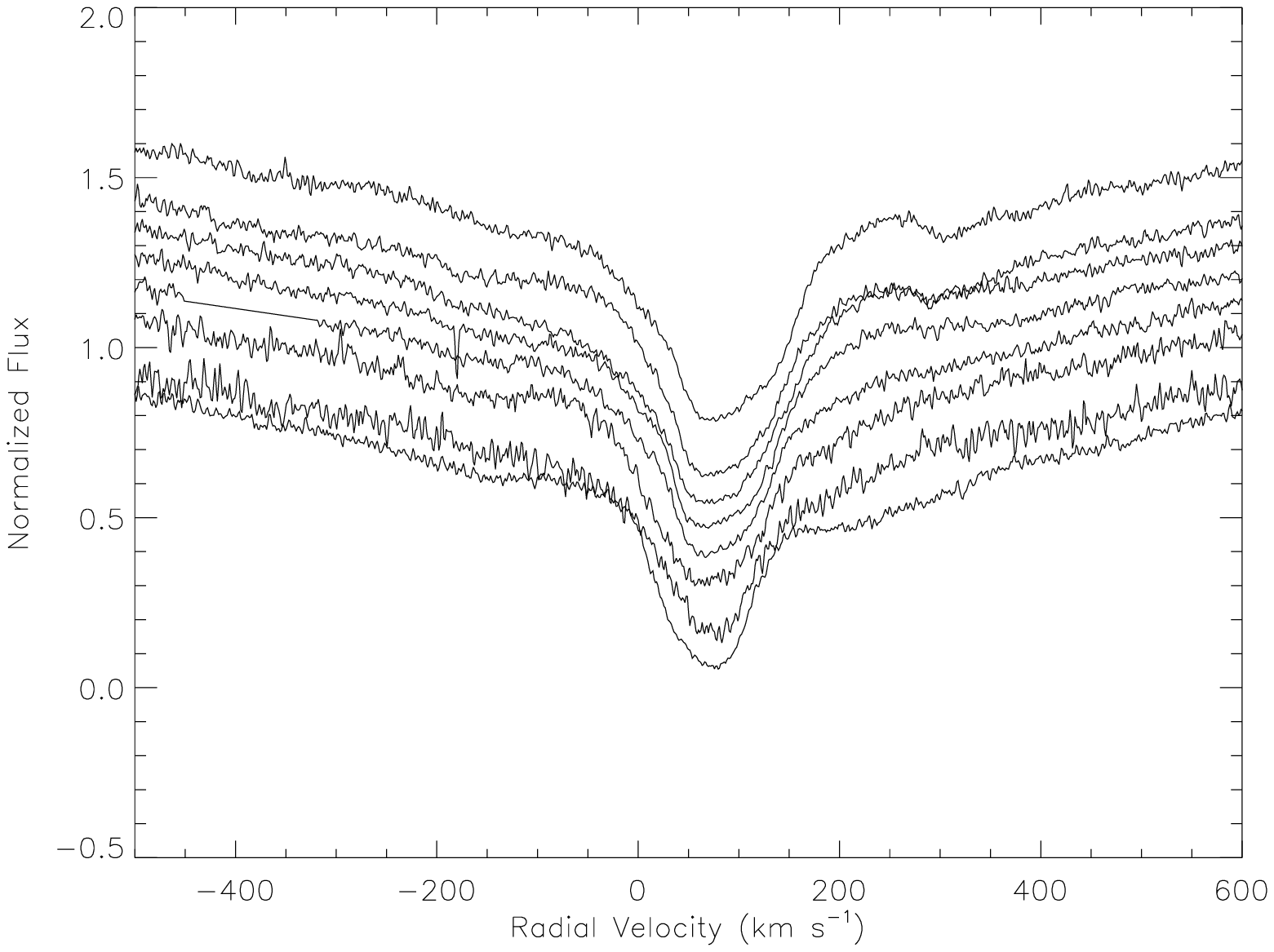}\qquad
\includegraphics[width=0.48\hsize]{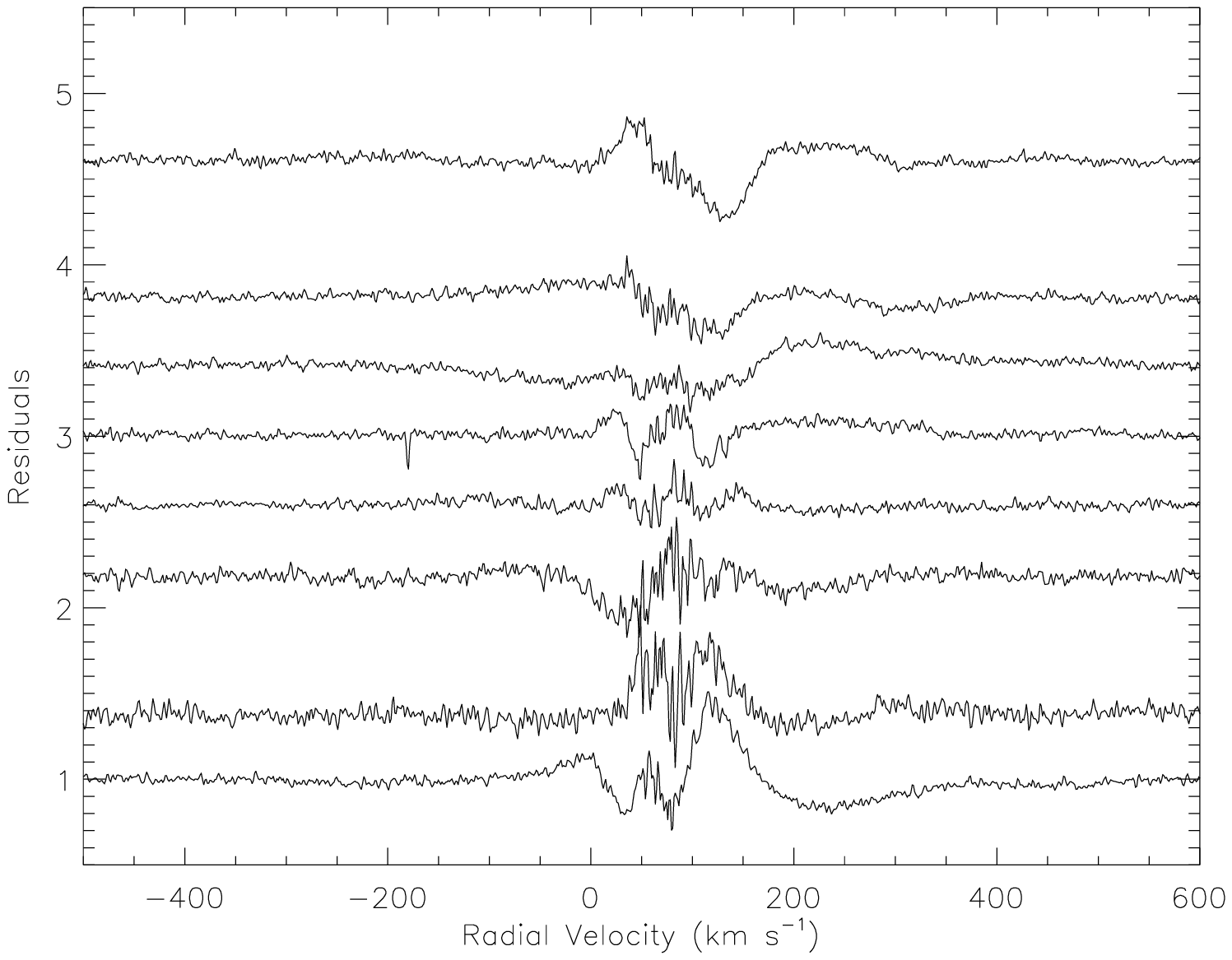}\qquad 
\caption{As in Fig.\,1, but for H\,9\,$\lambda 3835$ line.}
\label{H9}
\end{figure*}
}

\subsection{Rotational velocity}\label{rotvelo}

The stellar rotational velocity projected to the line of sight ($\varv\sin i$) 
can be determined from photospheric lines that are neither blended nor polluted 
by emission. As mentioned previously, the spectrum of HD\,50138 shows only three purely photospheric lines: He\,{\sc i}\,$\lambda 
4026$ and Si\,{\sc ii}\,$\lambda \lambda 4128, 4131$. However, the obvious
large line width results in blending of the Si\,{\sc ii} lines so that we 
are left with a single line, the He\,{\sc i}\,$\lambda 4026$ line. It can thus be regarded as the only feasible 
photospheric line. 

\begin{figure}[t!] 
\resizebox{\hsize}{!}{\includegraphics{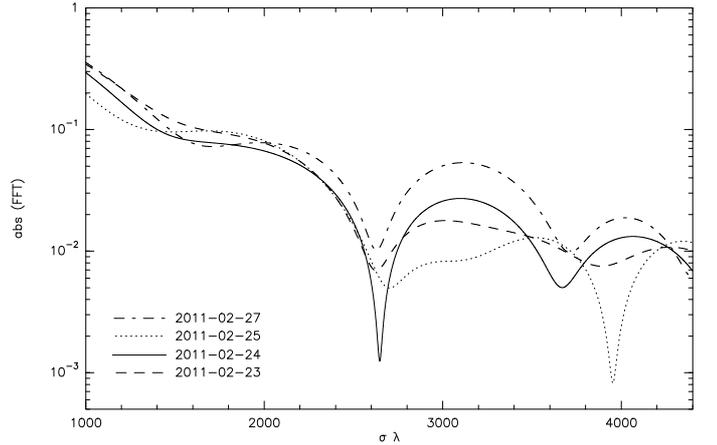}}
\caption{Position of the first zero point in the Fourier transform of
the line profiles of the He\,{\sc i}\,$\lambda 4026$ line.}
\label{vsini}
\end{figure}

To determine $\varv\sin i$ we applied the well established Fourier method 
(e.g., Gray \cite{Gray}, Sim\'on-D\'iaz \& Herrero \cite{SimonDiaz}). The 
Fourier transform of the stellar rotation profile possesses zero points, and 
the first zero is inversely proportional to $\varv\sin i$. To obtain reasonable 
results, the spectra are required to have both a high spectral resolution and 
a high $S/N$. While the HERMES spectrograph provides an excellent resolution with
$R\sim 85\,000$, some of our data suffer from poor S/N. 
We therefore restricted the analysis to four spectra with reasonable S/N 
values. The results are presented in Fig.\,\ref{vsini}. The agreement in the 
position of the first zero point in the Fourier transform of these lines is 
excellent. The $\varv\sin i$ values obtained from these zero points are listed 
in Table\,\ref{velocities}, and the resulting mean value is $\varv\sin i= 74.7
\pm 0.8$\,km\,s$^{-1}$.

\begin{table}[t!]
\caption{Rotational velocities ($\varv\sin i$) derived from the 
He{\sc i}\,$\lambda 4026$ line.}
\label{velocities}
\begin{center}
\begin{tabular}{ccc}
    \hline
    \hline
Date & $\varv\sin i$ & S/N \\
 & [km\,s$^{-1}$] & \\
    \hline
2011-02-23 & 75.39 & 240  \\
2011-02-24 & 74.75 & 230  \\
2011-02-25 & 73.36 & 230  \\
2011-02-27 & 75.30 & 220  \\
    \hline
\end{tabular}
\end{center}
\end{table}

Earlier attempts to derive the projected stellar rotational velocity of
HD\,50138 have been made by Corporon \& Lagrange (\cite{Corporon}) and
Fr\'{e}mat et al. (\cite{Fremat}), who obtained values for $\varv\sin i$ of
$50\pm 30$ and $60\pm 10$\,km\,s$^{-1}$, respectively. Considering the large
error bars obtained by these authors, their values are in reasonably good
agreement with our value. Corporon \& Lagrange (\cite{Corporon}) measured their
value from the He\,{\sc i}\,$\lambda\lambda$4472, 6678 lines, which are both
highly variable in our spectra (for He\,{\sc i}\,$\lambda 6678$ line see Fig.\,\ref{HeI6678}). Also Fr\'{e}mat et al. (\cite{Fremat}) used
highly variable lines like the He\,{\sc i} and Mg\,{\sc ii} lines at $\lambda
\lambda$4472, 4481, and their $S/N$ was considerably lower
than ours with about 150. Our value, which is obtained from the He\,{\sc i}\,$\lambda 4026$ line, the stablest line in the whole spectrum, can thus be regarded as reliable.

Considering in addition an inclination of $i = 56\degr \pm 4\degr$ of the star as
obtained for its dusty disk from an analysis of interferometric observations in Paper\,II, the stellar rotational
velocity of HD\,50138 becomes $\varv_{\rm rot} = 90.3\pm 4.3$\,km\,s$^{-1}$.
Applying the stellar parameters obtained in Paper\,I, the rotational period of HD\,50138 is $P= 3.64\pm 1.16$\,d.

\begin{figure}[t!]
\resizebox{\hsize}{!}{\includegraphics{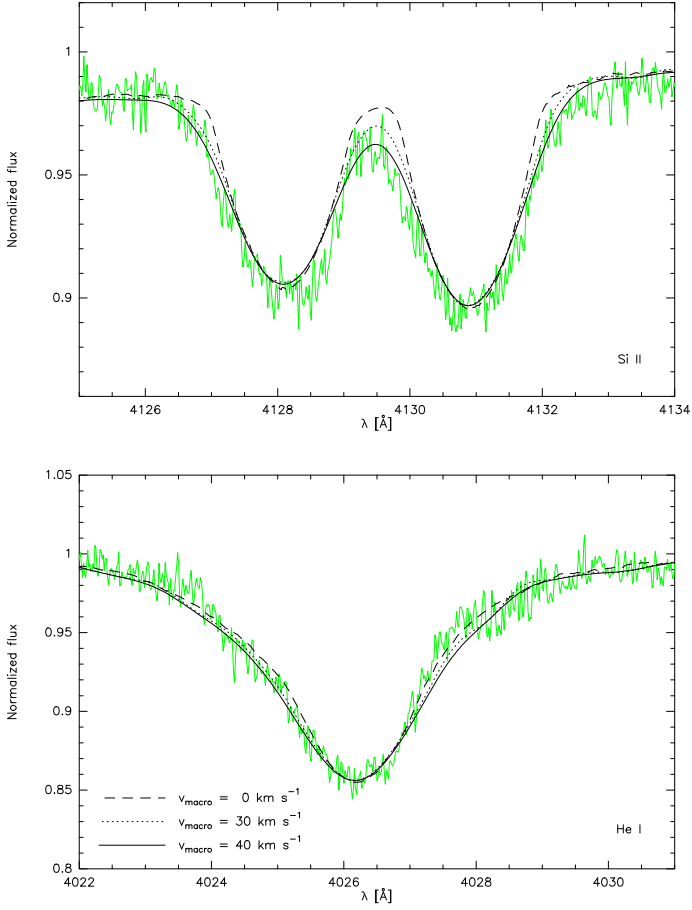}}
\caption{Fit of the observed Si\,{\sc ii} (top) and He\,{\sc i} (bottom) lines
with synthetic photospheric lines, using stellar parameters in agreement with Paper\,I, broadened with stellar rotation and macroturbulence.}
\label{linefits}
\end{figure}

Next we computed synthetic, rotationally broadened photospheric line profiles 
for stellar parameters of $T_{\rm eff} = 13\,000$\,K and $\log g = 4.0$, in agreement with Paper\,I, which
we fitted to the observed He\,{\sc i} and Si\,{\sc ii} lines, seen in the combined spectrum of 2011-02-27. These are shown as the dashed lines in Fig.\,\ref{linefits}.  

Comparing the synthetic and observed profiles, it is obvious that stellar 
rotation alone cannot account for the observed line width. This is particularly
evident for the Si\,{\sc ii} lines (top panel of Fig.\,\ref{linefits}). To fit
the observed profiles it was necessary to broaden the lines with an 
additional, Gaussian shaped profile. The needed amount of the extra velocity
broadening was about 30-40\,km\,s$^{-1}$. Line profile fits using 
these two values are shown in Fig.\,\ref{linefits} as dotted and solid lines, 
respectively.

The presence of such an excess broadening was also reported for another group 
of stars, the B supergiants (e.g., Ryans et al. \cite{Ryans}; Sim\'on-D\'iaz 
\& Herrero \cite{SimonDiaz}; Markova \& Puls \cite{MarkovaPuls}). These stars 
are furthermore known to display strong line profile variability (e.g., Kaufer 
et al. \cite{Kaufer97, Kaufer2006}; Lefever et al. \cite{Lefever}; Markova et 
al. \cite{Markova}; Clark et al. \cite{Clark}), and recent studies by
Sim\'on-D\'iaz et al. (\cite{SimonDiaz2010}) have revealed that a correlation 
exists between line profile variability in B supergiants and their excess 
broadening, which is usually referred to as \lq macroturbulence\rq.

\subsection{Moment analysis}\label{momcalc}

While the origin of the macroturbulence velocity seen in the line widths of  
B supergiants was long unclear, recent studies by Aerts et al. 
(\cite{Aerts2009}) suggest that the broadening might result from many 
(hundreds) individual contributions provided by stellar pulsations.
%
%
If pulsations could be a plausible explanation for the excess broadening seen 
in the He\,{\sc i} line of the non-supergiant B[e] star HD\,50138, then this line 
should also show profile variability to at least some extent. Since the 
residuals of this line do not reveal any obvious variability, a more thorough 
analysis of its profile is needed to unveil any small-scale (periodical) 
variability.

To investigate the putatively stable He\,{\sc i} line in more detail, we make use 
of the moment method, which is an excellent tool for studying even small-scale line 
profile variabilities and distinguishing between different physical mechanisms 
as the origin of the variabilities, such as (non-)radial pulsations and stellar 
spots. But for a reliable distinction between the two scenarios (pulsations 
versus chemical surface inhomogeneities in the form of spots), the variability 
in different chemical elements needs to be investigated. Besides the 
He\,{\sc i}\,$\lambda 4026$ line, we therefore included the Si\,{\sc 
ii}\,$\lambda\lambda 4128, 4131$ lines in the analysis, but keeping in mind
that the results of these lines probably have a higher uncertainty because
of their blending. We followed the 
description of Aerts et al. (\cite{Aerts}) and North \& Paltani (\cite{North}) 
by computing the equivalent width, together with the first three moments of 
these lines. Due to their poor quality, the spectra taken on 2011-02-20 and 
2011-02-21 were excluded. 

\begin{figure*}[t!]
\resizebox{\hsize}{!}{\includegraphics[angle=270]{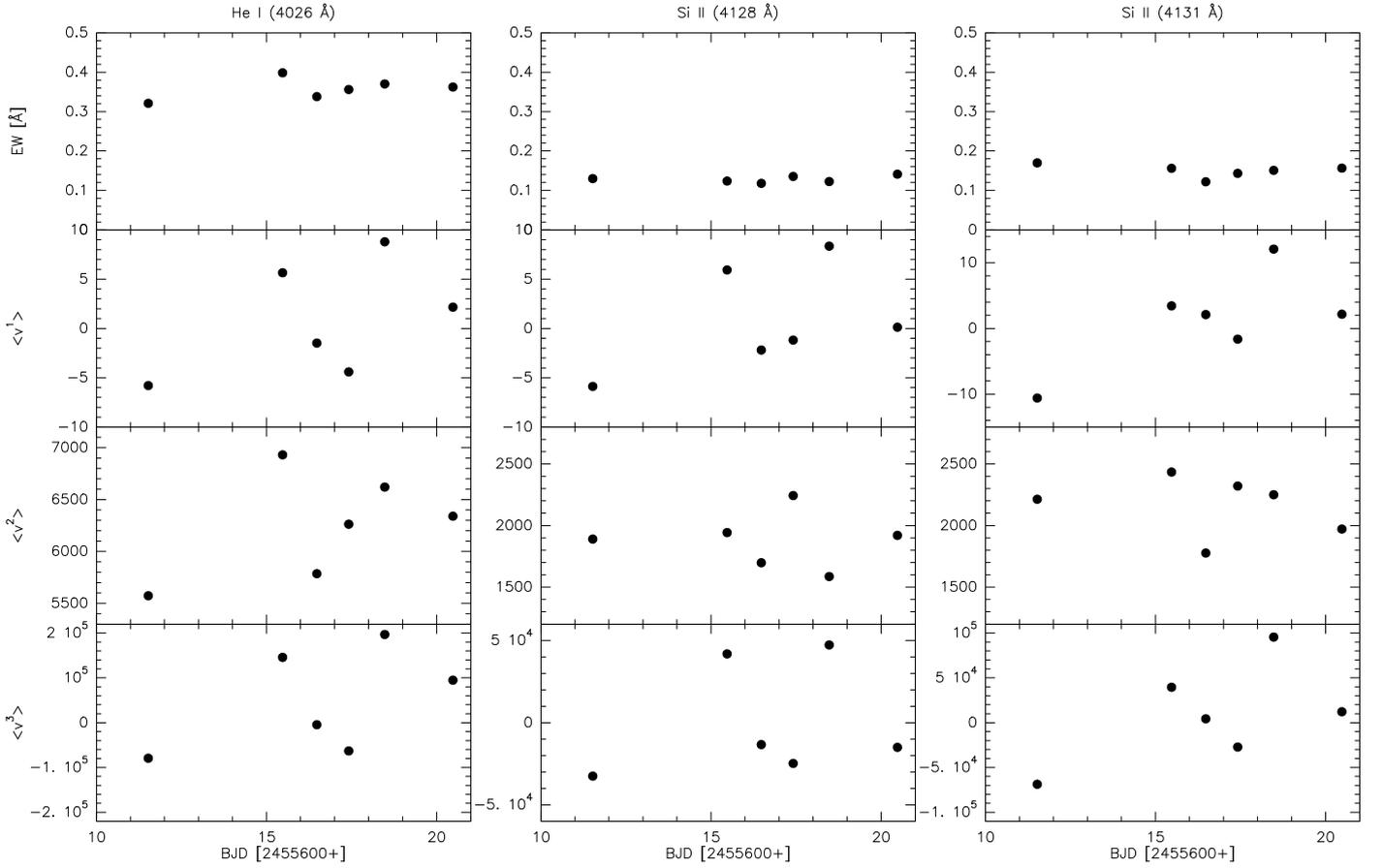}}
\caption{Variation in the equivalent widths and first three moments calculated for the He\,{\sc i}\,$\lambda 4026$ and Si\,{\sc ii}\,$\lambda\lambda 4128, 4131$ with time. The moments $<\varv^{1}>$, $<\varv^{2}>$, and $<\varv^{3}>$, have units (km\,s$^{-1}$), (km\,s$^{-1}$)$^{2}$, and (km\,s$^{-1}$)$^{3}$, respectively.}
\label{moments}
\end{figure*}

Figure\,\ref{moments} shows the resulting time variation of the equivalent widths 
and the first three moments, which represent the radial velocity 
($<\varv^{1}>$), a measure of the width ($<\varv^{2}>$), and the skewness 
($<\varv^{3}>$) of the line. Quite clearly, all three lines show variability in their first three moments, while their equivalent widths (zeroeth moment) remain relatively stable. The amplitude in radial velocity variation is on the order of
10\,km\,s$^{-1}$. It is slightly higher in the Si\,{\sc ii} lines. This might
explain why the residuals in those lines show some signal (see Fig.\,\ref{SiII4128}), while those
of the He\,{\sc i}\,$\lambda 4026$ line are almost featureless. 

\section{Discussion}\label{discussion}

The very similar variation in both He\,{\sc i} and Si\,{\sc ii} lines makes
it unlikely that their variabilities could originate in a star with a
heterogeneous surface abundance pattern in the form of stellar spots. Such stars 
usually show a quite different behavior in the variability seen in their 
helium lines with respect to those in their metal lines (see, e.g., Briquet et 
al. \cite{Briquet01, Briquet04}; Lehmann et al. \cite{Lehmann}). 

In addition, the line profiles in spotted stars vary with their rotational
period. Owing to an insufficient time coverage of our observations (only one 
data point per night), a search for periodicity in the moments failed. However, the rapid change in radial velocity from -5 to about +10\,km\,s$^{-1}$ in the He{\sc i} line within less than 26\,hours (from BJD 17.416 to 18.473), while it showed a steady decrease from $\sim 6$ to $\sim -5$\,km\,s$^{-1}$ during the three preceding nights,
might indicate (using a tentative sine curve fit) a possible (maximum) period of $\sim$ 1.7 days, but the real period could also be (much) shorter. This needs to be studied on time series of spectroscopic observations taken with a much better time resolution. 

\begin{figure}
\resizebox{\hsize}{!}{\includegraphics{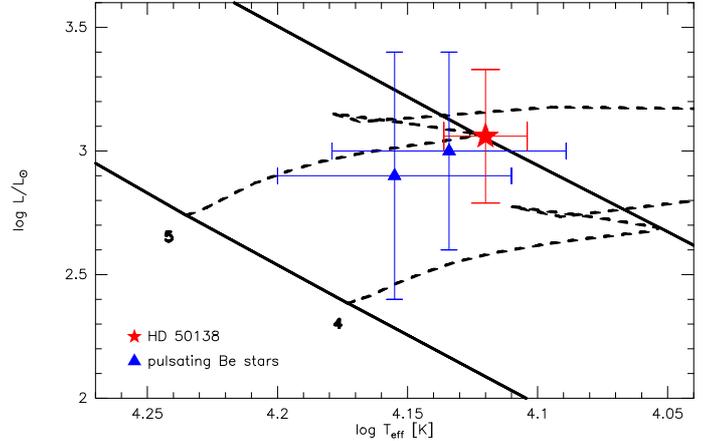}}
\caption{Hertzsprung-Russell diagram showing two pulsating late-type Be
stars in comparison with HD\,50138. Parameters of the Be stars are from
Neiner et al. (\cite{Neiner09}) and Diago et al. (\cite{Diago09}). Solid lines encompass the main sequence, dashed lines represent the evolutionary tracks for stars with M$_{ZAMS}$ = 4 and 5 M$_\odot$.}
\label{hrd_Be}
\end{figure}

Consequently, the variability in the lines of HD\,50138 is also not connected with the rotation period of $\sim$ 3.6\,days. Instead, it resembles those seen in pulsating stars. In addition, the strong short-term variability of the line profiles is mainly contained in the lines formed in the upper layers of the stellar atmosphere or very close to the stellar surface (like the core of the higher Balmer lines), while the lines formed in the circumstellar environment (like the [O\,{\sc i}] lines) are not affected by these night-to-night variations. It is interesting to point out that two pulsating late-type Be stars exist and occupy a similar region in the Hertzsprung-Russell diagram as our object (see Fig.\,\ref{hrd_Be}). 

In addition, the fact that macroturbulence is present in the photospheric lines of HD\,50138, a known property of B supergiants, and its classification as a star at the end of (or slightly beyond) the main sequence, makes it a fascinating object.

\section{Conclusions}\label{conclusion}

Based on new high-resolution spectra of HD\,50138 obtained during eight nights (five of them consecutive), we confirm the presence of short-term variability in a large sample of lines of different elements. These spectral variations have different strengths depending on the line formation region. The strongest short-term variability is identified in the profiles of lines arising in the upper layers of the stellar atmosphere or very close to the stellar surface. 

Analyzing the few present photospheric lines, we derived the rotational velocity and rotational period of HD\,50138, $\varv_{\rm rot} = 90.3\pm 4.3$\,km\,s$^{-1}$ and $P= 3.64\pm 1.16$\,d, respectively. We conclude that the short-term line profile variability is not connected with this rotation period. In addition, we used the moment method to study the origin of the line profile variability and discard any possibility that there are stellar spots, which suggest pulsations as the origin of the short-term variations identified. However, from our data it was not possible to determine the modes of these pulsations. Nevertheless, the scenario of pulsations, though putative at that moment, might also favor the appearance of shell phases as reported in Paper\,I. As in pulsating Be stars, these mass ejections could be associated to possible beat periods.

These new results, accompanied by the probable macroturbulence in the photospheric lines, reveal the fascinating and complex properties of HD\,50138, an object clearly in need of more detailed study. An observational campaign to obtain more data of higher quality, in a longer temporal series, is definitely necessary. From these data, it will be possible to precisely determine the real origin of the line profile variations and consequently the nature of this star. 

\begin{acknowledgements}
This research made use of the NASA Astrophysics Data System (ADS). The authors acknowledge the referee, Dr. Mikhail Pogodin, for his useful comments, which allowed us to improve our manuscript. MBF acknowledges Conselho Nacional de Desenvolvimento Cient\'{\i}fico e Tecnol\'ogico (CNPq-Brazil) for the post-doctoral grant. M.K., D.H.N., and M.E.O. acknowledge financial support from GA\v{C}R under grant number P209/11/1198. The Astronomical Institute Ond\v{r}ejov is supported by the project RVO:67985815.
\end{acknowledgements}

\end{document}